\documentclass[conference]{IEEEtran}
\IEEEoverridecommandlockouts

\usepackage{todonotes}
\usepackage{changes}
\definechangesauthor[name={ilya}, color=red]{is}
\definechangesauthor[name={Farah}, color=green]{fa} 

\definechangesauthor[name={Alex}, color=orange]{ah}

\usepackage[bookmarks=false]{hyperref}

 \hypersetup{
    colorlinks=true,
    linkcolor=black,
    urlcolor=blue,
    citecolor=black,
}

\usepackage{cite}
\usepackage{amsmath,amssymb,amsfonts}
\usepackage{algorithmic}
\usepackage{graphicx}
\usepackage{textcomp}
\usepackage{xcolor}
\def\BibTeX{{\rm B\kern-.05em{\sc i\kern-.025em b}\kern-.08em
    T\kern-.1667em\lower.7ex\hbox{E}\kern-.125emX}}
\begin{document}


\title{Accelerating Text Mining Using  \\ Domain-Specific Stop Word Lists
}

\author{
\IEEEauthorblockN{Farah Alshanik, Amy Apon, Alexander Herzog, Ilya Safro, Justin Sybrandt 
}
\IEEEauthorblockA{\textit{School of Computing} \\
\textit{Clemson University}\\
Clemson, USA \\
\{falshan, aapon, aherzog, isafro, jsybran\}@clemson.edu}
\vspace{-.4in}
}


\maketitle

\begin{abstract}
Text preprocessing is an essential step in text mining. Removing words that can negatively impact the quality of prediction algorithms or are not informative enough is a crucial storage-saving technique in text indexing and results in improved computational efficiency. 
Typically, a generic stop word list is applied to a dataset regardless of the domain. However, many
common words are different from one domain to another but have no significance within a particular domain. Eliminating domain-specific common words in a corpus  reduces  the dimensionality of the feature space, and improves the performance of text mining tasks. 
In this paper, we present a novel mathematical approach for the automatic extraction of domain-specific words called the hyperplane-based approach. 
This new approach depends on the notion of low dimensional representation of the word in vector space and its distance from hyperplane. 
The hyperplane-based approach can significantly reduce text dimensionality by eliminating irrelevant features. 
We compare the hyperplane-based approach with other feature selection methods, namely $\chi^2$ and mutual information. 
An experimental study is performed on  three different datasets and five classification algorithms, and measure the dimensionality reduction and the increase in the classification performance. 
Results indicate that the hyperplane-based approach can reduce the dimensionality of the corpus by 90\% and outperforms mutual information.
The computational time to identify the domain-specific words is significantly lower than mutual information. 

Reproducibility: code and results can be found
at \url{https://github.com/FarahAlshanik/Domain-Specific-Word-List}





\end{abstract}





\section{Introduction}

The amount of raw text data produced in science, finance, social media, and medicine is growing at an unprecedented pace. Without effective preprocessing, the raw text data typically introduces major computational and analytical obstacles (e.g., extremely high dimensionality) to data mining and machine learning algorithms. Hence, text preprocessing 
has become an essential step to better manage and handle various challenges inherent with the raw data before it is ready for text mining tasks. 
Stop word removal is one of the most important steps in text preprocessing. 
Stop words are those that appear frequently and commonly and contribute little analytical meaning and value in text mining tasks\cite{zou2006automatic,chong2012empirical}.  

Elimination of stop words reduces text count in the corpus typically by 35\% to 45\%  
and makes the application of text mining methods more effective\cite{jha2016hsra}. 
In the  majority of text mining tasks,  preprocessing begins with eliminating stop words without considering the domain of application.
For example, in the popular Python Natural Language
ToolKit (nltk) module, the English stop words list contains  frequent words such as ``the'', ``and'', ``to'', ``a'', 
and ``is'' \cite{bell1990text}. 
However, there may exist many  domain-specific words that do not add value in a text mining application but that increase the computational time. 
For instance, the word “protein” is likely to add little value in  text mining of  bioinformatics articles, but not for a collection of political articles. Reduction in running time is also extremely important for such tasks as  streaming and sliding window text mining \cite{avudaiappan2017detecting,9005964}.

The objective of this paper is  to  present the hyperplane-based approach for detecting and eliminating  domain-specific common words in a corpus of documents in addition to (or in some cases instead of)  commonly-used   stop   words. Our approach aims to enhance
the performance of binary text classification and to reduce the
dimensionality. In  binary text classification, the documents
are grouped into two classes. The trained classifier’s role is to predict
the document class. However, the domain-specific words often mislead
the classifiers’ prediction. The  hyperplane-based approach employs the principle of low dimensional representation of words
using the word2vec skip-gram model. The main idea of our approach is to project
in the vector space both centroids of the two classes and each word in the corpus. The method  constructs a hyperplane perpendicular to the normalized vector between the centroids of the two classes such that the words closer to the perpendicular hyperplane are selected as domain-specific common words.

To justify the proposed approach and to demonstrate the contextual difference with other traditional feature elimination methods, we eliminate a varying count of the detected domain-specific words from the corpus and observe the accuracy of five different classifiers: Naive Bayes, Logistic Regression, Random Forest, Support Vector Machine (SVM), and Classification and  Regression Trees (CART). 
We evaluate the five approaches on three different  datasets: (1) Pubmed \cite{website2}, (2) IMDB movie reviews\cite{maas-EtAl:2011:ACL-HLT2011}, and (3) Hansard Speeches 1979–2018\cite{odell2019}.


Two major experiments were performed.
First, we observed the performance of our approach before and after eliminating domain-specific words on the classification accuracy and execution time.
Secondly, we compared the classifier accuracy and validate our results with other feature selection methods namely $\chi^2$ and mutual information. The classifiers are used to predict the journal name for the Pubmed dataset, sentiment for the IMDB movie review dataset, and party for the Hansard  Speeches. Also, we found the size of the overlap of the remaining words after eliminating the domain-specific words between our approach and the $\chi^2$ or mutual information.

The results show that the hyperplace-based approach has improved performance in terms of accuracy and execution time with respect to feature selection methods. Through our proposed approach we identify that a significant number of stopwords can be removed without reducing the accuracy of any classifier we considered. On the contrary, we determine that after removing 90\% of the words within the Hansard Speeches dataset according to our method, the accuracy of a Random Forest classifier increases by 4\%, and a Naive Bayes classifier increases by 12\%.



\section{Related Work}

Stop word removal  has been an active area of research
and several studies have sought an automated method for generating a stop word list for different languages. 
Some  studies try to find a generic stop word list that is domain independent. Others focus on finding domain-specific  words that are document-dependent\cite{kaur2018systematic}.
One method, term frequency, was the first automated process for identifying and extracting stop words.
The term frequency approach produces valuable results for the words that have the  highest frequency in the corpus, 
but term frequency does not take into account the words that occur frequently in only some documents. Also, the term frequency does not consider the words that rarely occur in the corpus but still meaningless for the classification.
\cite{1241221}. 
Our work differs from this approach by returning the domain-specific common words with respect to the document collection. 
 
In \cite{garg2014effect} the authors used term frequency for word elimination and studied its effect on the similarity of Hindi text documents. They found that the removal of stop words based on term frequency decreased  the similarity of documents.
In \cite{gupta2011preprocessing}
authors used term frequency with Punjabi texts. 
They identified 615 stop words in a corpus of more than 11 million words. However, in later work they found that the frequency count cannot  be taken as the true measure of stop word identification \cite{puri2013automated}. 
To overcome the drawback of the frequency words the authors used a statistical model based on the distribution of words in documents along with the frequency count to generate an aggregated stop word list for the Punjabi language. 

Word entropy is another method.
It was used in \cite{1241221} to generate web-specific stop lists for web document analysis.
This research aimed to eliminate common web-specific terms such as ’email’, ’contact’, etc.  
The words with the lowest entropy were considered to be stop words. 
The method was evaluated using the web document dataset and the BankSearch dataset. The authors successfully extracted the web-specific stop list from the web document dataset. 
However, the results on the BankSearch dataset, which consists of 10 different categories, showed a bias towards low entropies for words that are category-related and frequent in a large number of documents. The authors
claimed that to use the word entropy the dataset should be  unbiased towards any subject category. 
In \cite{zheng2011comparative} the authors used word entropy to generate a Mongolian stopword list.


Other research has combined the term frequency and word entropy approaches.
In \cite{rani2018automatic} the authors used strategies focused on statistical methods and knowledge-based measures for creating
generic stop words for Hindi text.
The objective is to measure the information content of each word in the corpus, which is measured using word entropy. 
The main advantage of this method is to overcome the problem inherited by manually picking stop words, which takes a lot of time. 
Authors in \cite{miretie2018automatic} aggregate term frequency and entropy value to construct the first stop word list for the Amharic text. 

A Chinese corpus consisting of Xinhua News and People’s Daily News is used in \cite{zou2006stop} to find a generic stop word list aggregated from two models:  a statistical model that
measures the combination
of mean probability and variance of probability, and an information model
that measures the informativity of words using the words’
entropy in the corpus. The generated Chinese stop word list
had a high intersection with the English stop words. Also, compared with other Chinese stop lists, this
method was more effective, and it was faster than  manual generation\cite{zou2006stop}.
The methodology was used to generate an  Arabic stop word list in \cite{alajmi2012toward} with superior performance. 

Authors in \cite{lo2005automatically} present the term-based random
sampling approach in which they generated a list of stop words
based on the importance of terms using Kullback-Leibler
divergence measure, which models how informative
a term is. They show that term-based
random sampling outperforms the rank-frequency
approach in terms of computational effort to
derive the stop word list. However,  using the term-based random sampling approach on  DOTGOV and the WT10G collections to obtain the stop word list did not provide  better results than baseline approaches.

A dictionary approach is used in \cite{raulji2016stop} to generate the first Sanskrit stop word
list. 
However, this method of generating the stop word list is both resource intensive and time consuming\cite{miretie2018automatic}. 
The hyperplane-based approach does not require a dictionary to return the domain-specific words.
A rule-based
approach using static eleven rules was used in \cite{rakholia2017rule} to automatically develop a stop word
list for the Gujarati language. 
The main limitation of this approach, as described in the paper, appears when the stop words contain more than three characters. 
In the hyperplane-based approach there is no limitation based on the length of the domain-specific word.

A method based on the weighted 
$\chi^2$ was used to generate the Chinese stop word list\cite{4721850}.
The method considers the high document frequency 
and the low statistical correlations with all the classification
categories. 
This is the first list that uses the dependent relationship between a word and all categories in a set of documents. 
This approach inputs a threshold to specify the size of the stop word list.  
The Naive Bayes classifier tested the performance of the generated stop words before and after elimination, and tested stop word lists of various lengths. 
The study eliminated the words based on some threshold. 
However, there is no guidance on the optimal elimination percentages of the detected domain-specific words. 
In our work we have performed extensive experiments to find and provide guidance on the best percentage of elimination.

\section{Background}


The low-dimensional representation method, word2vec, is utilizing the skip-gram model. The word2vec maps words into a low-dimensional space  
revealing non-trivial context based relationships between them. Many syntactic and semantic relationships between words can be defined by using simple algebraic operations on the word vectors.  For example, the word vector "King" - word vector "Man" + word vector "Woman" results in a word vector similar to vector representation of the word vector "Queen"\cite{mikolov2013linguistic}.

Naive Bayes classifier\cite{lewis1998naive}, 
a probabilistic model based on Bayes theorem, is a supervised learning technique broadly used for 
text clasification. 
The method is, in practice, extremely sensitive to preprocessed data quality and other factors. Here we mostly use it as a baseline technique to reflect the sensitivity of the preprocessing.

Support Vector Machine (SVM) is a quadratic  optimization-based technique that, in its simplest form, maximizes the margin between classes using the optimized separating hyperplane. While having fast performance, the linear SVM is known to be sensitive to  hard classification problems. Slower nonlinear (aka kernel based) SVM often produces higher quality results but requires suboptimal heuristics to achieve linearly scalability. \cite{sadrfaridpour2019engineering}. Here we use the Thunder SVM package \cite{wenthundersvm18} which provides a good quality/performance trade-off.


We also use three other broadly applicable  classification techniques, namely, Logistic Regression\cite{lever2016logistic}, Random Forest\cite{ho1995random}, and Classification and Regression Trees (CART)\cite{breiman1984classification}.
These are methods of comparable quality when applied in the text mining domain. Their implementation is available inside Python's module called Scikit-learn\cite{scikit-learn}.




\section{Domain-specific word detection}

\subsection{Simplified Example}
\begin{figure}
   \includegraphics[width=0.7\linewidth,height=4cm]{ 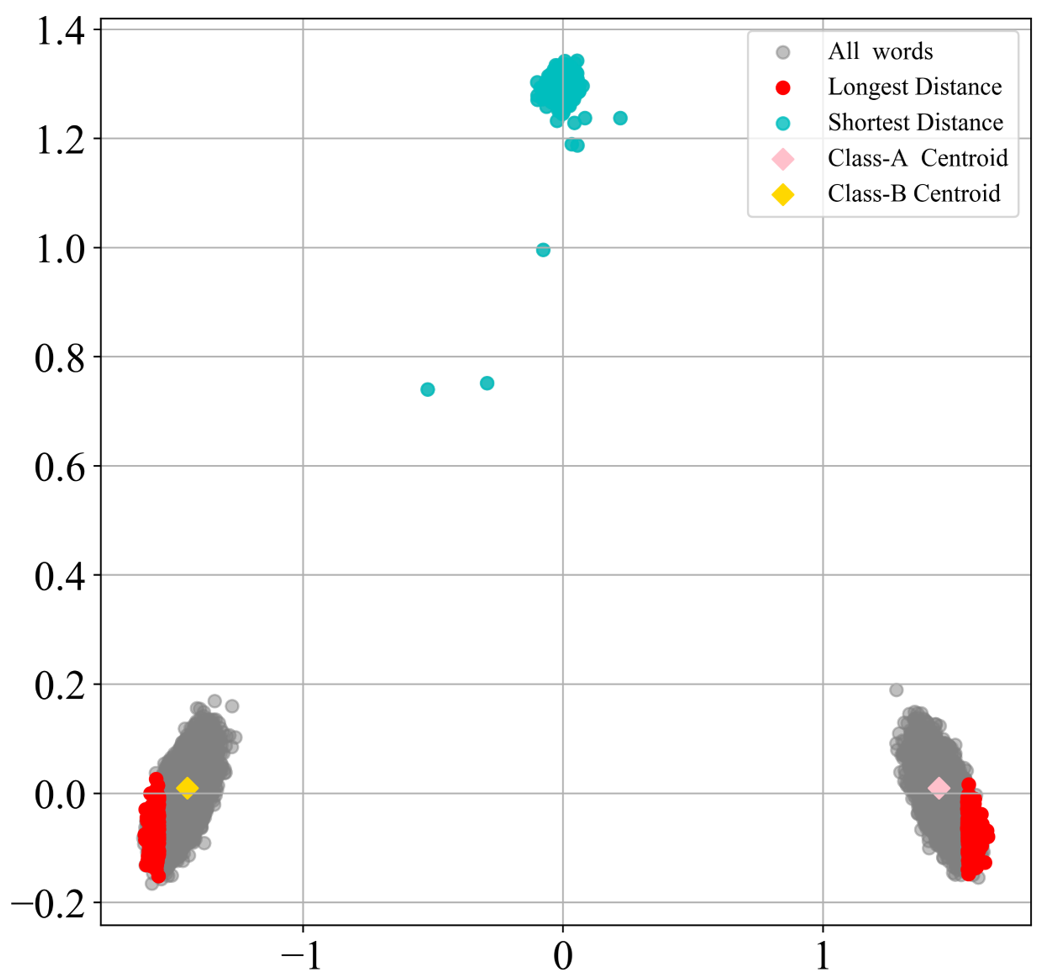}
   \centering
   \vspace{-.1in}
   \caption{PCA-based visualization of the synthetic dataset embedding. The closest words to hyperplane are
marked in blue (300 words). The words that have a largest distance are marked in red (300 words), the centroid embedding of the class $A$ and class $B$ are marked in pink and gold,
respectively}
\vspace{-.2in}
   \label{synthetic_data}
   \end{figure}

The basics of our approach are best understood with a simple example on a synthetic dataset.  
Suppose a synthetic binary class dataset of total of 40,000 documents is generated in which the two classes have an equal number of documents. 
The length of each document from both classes is 300 words. The words are chosen randomly from predefined dictionary that has been created for this experiment. 
Dictionaries of classes $A$ and $B$ consist of 2000 words each represented by $w_i$, and $v_i$, respectively, where $ 1\leq i \leq 2000$. The classes are disjoint and any classifier will result in a 100\% accuracy. Now, in order to include common words between the classes, a dictionary of 300 words $m_i$ is created, $1\leq i\leq 300$. 
Then, 10 $m_i$ words are randomly added to each document in the two class. The word2vec skip-gram is applied on the synthetic data corpus to project the words into a vector space using 100-dimensional word vectors.
After that the centroid of each class is calculated by averaging the word embedding in each class. We found that the common words $m_i$ exist in the middle between the two classes as shown in Fig.~\ref{synthetic_data} (blue points). Accordingly, we decided to create a perpendicular hyperplane on the normalized vector between the centroid of the two classes (pink and gold).  Additionally, we calculated the distance between each unique word in the corpus and the hyperplane, such that the words that have the shortest distance will be the domain-specific common words, and the words that have the longest distance will be the words that are used to distinguish between the two classes (red).

\subsection{Overview of the Approach}

We introduce the hyperplane-based approach to detect the domain-specific words.  It is based on the distance of the word from the separating hyperplane with the notion of a low dimensional vector representation of the words using the word2vec model\cite{mikolov2013efficient}. 
The domain-specific words are determined as the words that have the \emph{shortest distance} from the hyperplane. 

\begin{figure}

   \includegraphics[width=0.85\linewidth,height=8cm]{ 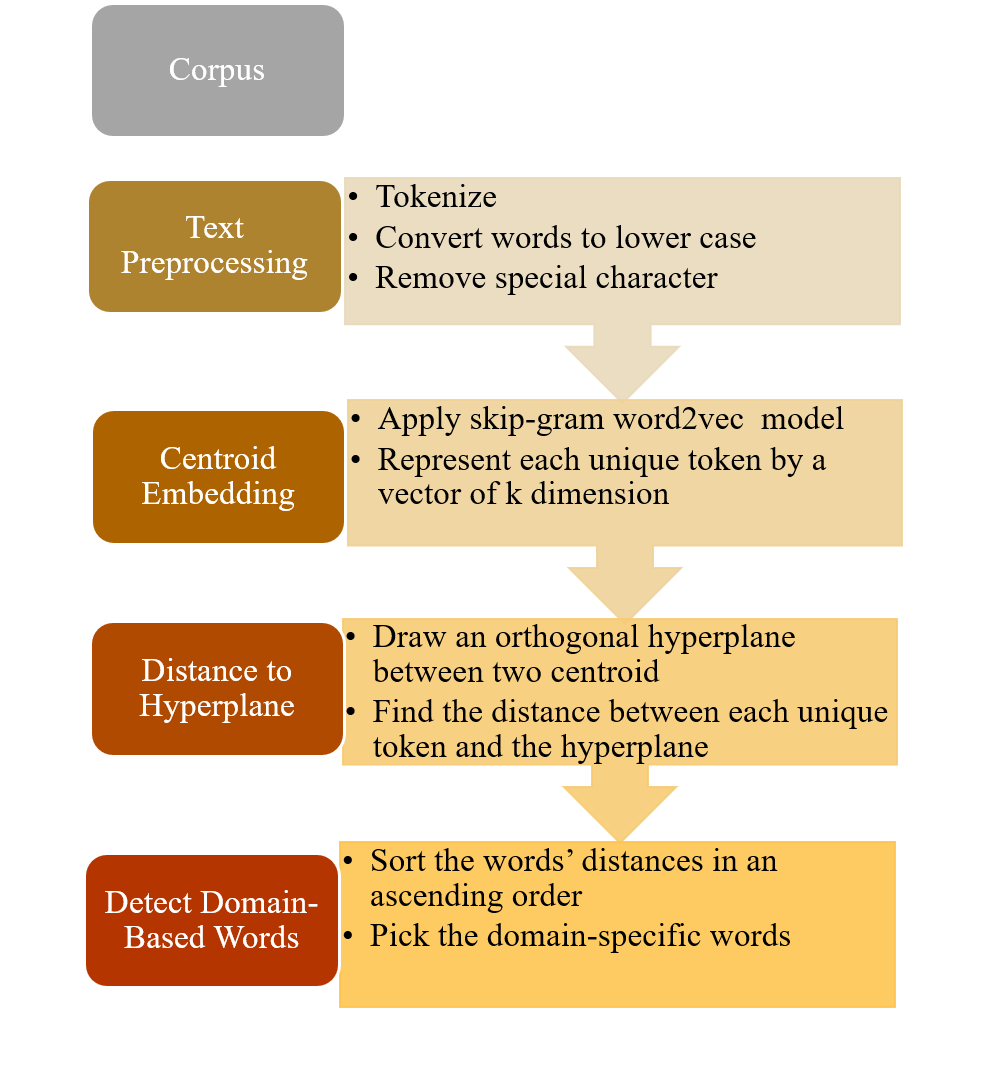}
   \centering
   \vspace{-.2in}
   \caption{Hyperplane-based approach steps }
   \label{flowchart}
 \vspace{-.2in}
   \end{figure}

This approach aims to enhance the performance of binary text classification and reduce the execution time of the classifier. The goal is to eliminate the features (i.e., words) that are not used to distinguish between the two classes. 
The flowchart of the approach is presented in Fig.~\ref{flowchart}.
The hyperplane-based method consists of four main steps: 1) text prepossessing, 2) centroid embedding,
3) computing the tokens' distances from an orthogonal hyperplane, and 4) sorting and detecting the domain-specific common words list. These steps are explained in the following subsections.

\vspace{-.05in}
\subsection{Text Preprocessing }

The text preproccesing step starts by tokenizing the text, then converting the words to lower case, and finally removing the special characters as shown in Fig.~\ref{flowchart}. Traditional stop words (such as those that are given in Python nltk module) are not removed.
Rather, some or all of the stop words are anticipated to be removed as a result of our method which indeed happens in the end of the entire process.




   

\subsection{Centroid Embedding }
The centroids of class are the averages over the corresponding word embeddings precomputed usign word2vec skip-gram  model. This results that the semantically similar words are mapped together in a vector space. This representation is also known as distributed numerical representations of word features. 

\begin{figure}
   \includegraphics[width=0.99\linewidth,height=8cm]{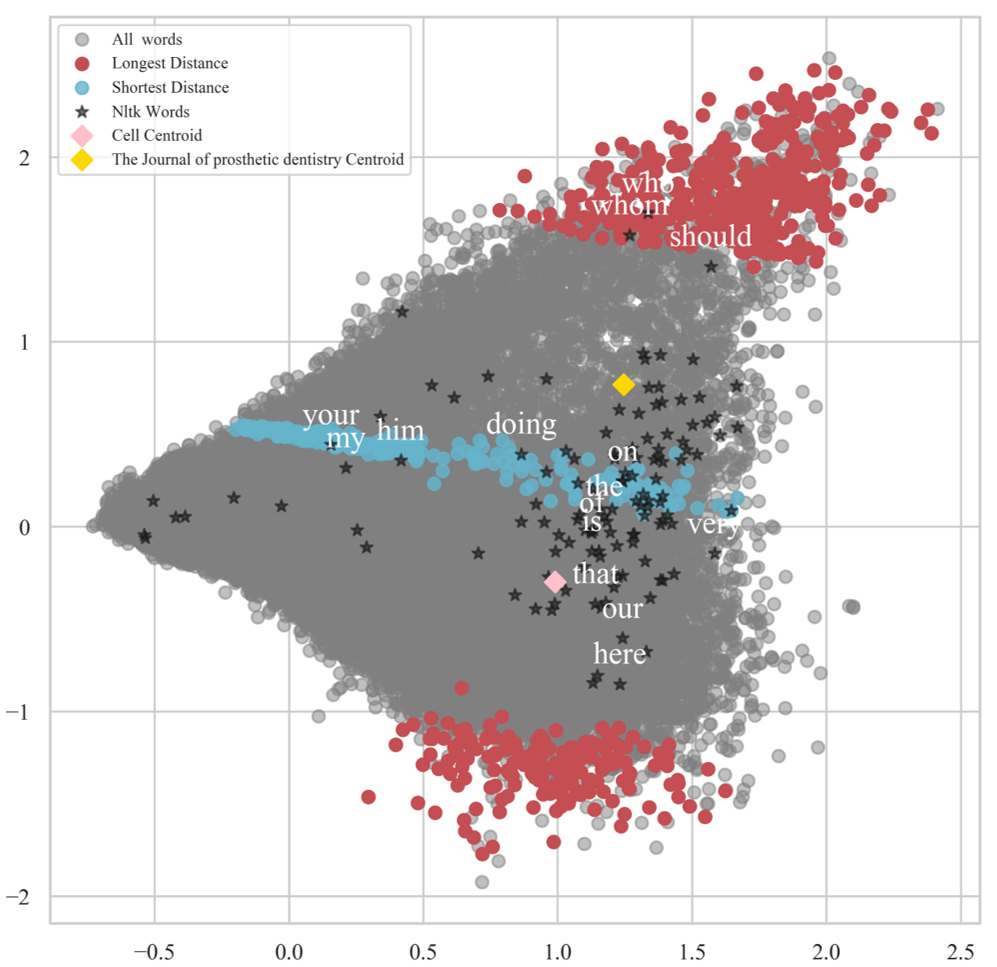}
   \vspace{-.1in}
   \caption{PCA-based visualization of the  Cell and Journal of Prosthetic Dentistry from Pubmed dataset. The closest words to hyperplane are
marked in blue (300 words), the words that have a largest distance are marked in red (300 words), the centroid embedding of the cell and Journal of Prosthetic Dentistry  journals are marked in pink and gold respectively, some nltk stop words are plotted in white. }
   \label{pca}
   \vspace{-.2in}
   \end{figure}

Given two disjoint classes of $n$ documents, namely, $A$ and $B$.
The centroid of a class is defined using the following steps. 
First, the word2vec skip-gram model is utilized to construct a $k$-dimensional representation of each word, 
%
i.e., each embedded word, or token, $t$ is represented as a vector $emb(t)=(t_{1},t_{2},t_{3},..., t_{k})$. 
%
%
%
 Then,  the centroid of each class is computed as the average of the embeddings of words from that class:
\begin{equation}
Center_X=\frac {\sum\limits_{t \in X} emb(t)}{M}\label{eq2},
\end{equation}
where $X$ is a class, $emb(t)$ is an embedding function, and $M$ is the total number of unique words in class $X$. 
In Fig.~\ref{pca}, we visualize the example of embedding using the 2D PCA dimensionality reduction from the initial 100-dimensional embedding space. In this example, the corpus of documents for classification contains abstracts of two journals, \emph{Cell} and \emph{Journal of Prosthetic Dentistry}, extracted from the Pubmed dataset. The figure illustrates the two-dimensional visualization of the individual journals' centroids and the stop words in nltk module, represented by diamonds and stars, respectively. 

\subsection{Distance from Hyperplane}
In this paper, the separating hyperplane non-strictly defines the boundary  between the two classes and their centroids, nanemly,  $A$ ($Center_A$)  and class $B$ $(Center_B)$. It is  orthogonal to the line connecting the centroids, and passes through its  middle. It is defined by the linear equation:

\begin{equation}
w'x+ b = 0\label{eq} 
\end{equation}

where $w$ is the slope of the plane, $b$ the offset. 
The first step is to find the slope of the plane, i.e. the normal vector perpendicular to the plane $(w)$, which is defined as the difference between the two centroids $A$ and $B$ as appear in equation \eqref{eq4}.

\begin{equation}
w=Center(A)-Center(B)\label{eq4}
\end{equation}

The offset point $(b)$ will be equal to the negative dot product of the normal vector by the coordinate of any point on the plane as expressed in equation \eqref{eq5}.
\begin{equation}
b =- w\cdot x_0\label{eq5}
\end{equation}


The predefined point at this stage is the mid point between the two centroids ($x_0$), which can be calculated using:

\begin{equation}
x_0=\frac{Center_A-Center_B}{2}
\end{equation}

After defining the hyperplane equation, the distance between any point (embedded words in space) to the hyperplane is calculated as: 

\begin{equation}
d= \frac{|w\cdot x_0+b|}{||w||} 
\end{equation}




Fig.~\ref{pca} shows that the shortest distance words (blue points) are clustered around the hyperplane which are the domain-specific words. The longest distance words (red points) are far away from the hyperplane which are the words that used to distinguish between the two classes.

\begin{figure}
   \includegraphics[width=0.99\linewidth,height=6cm]{ 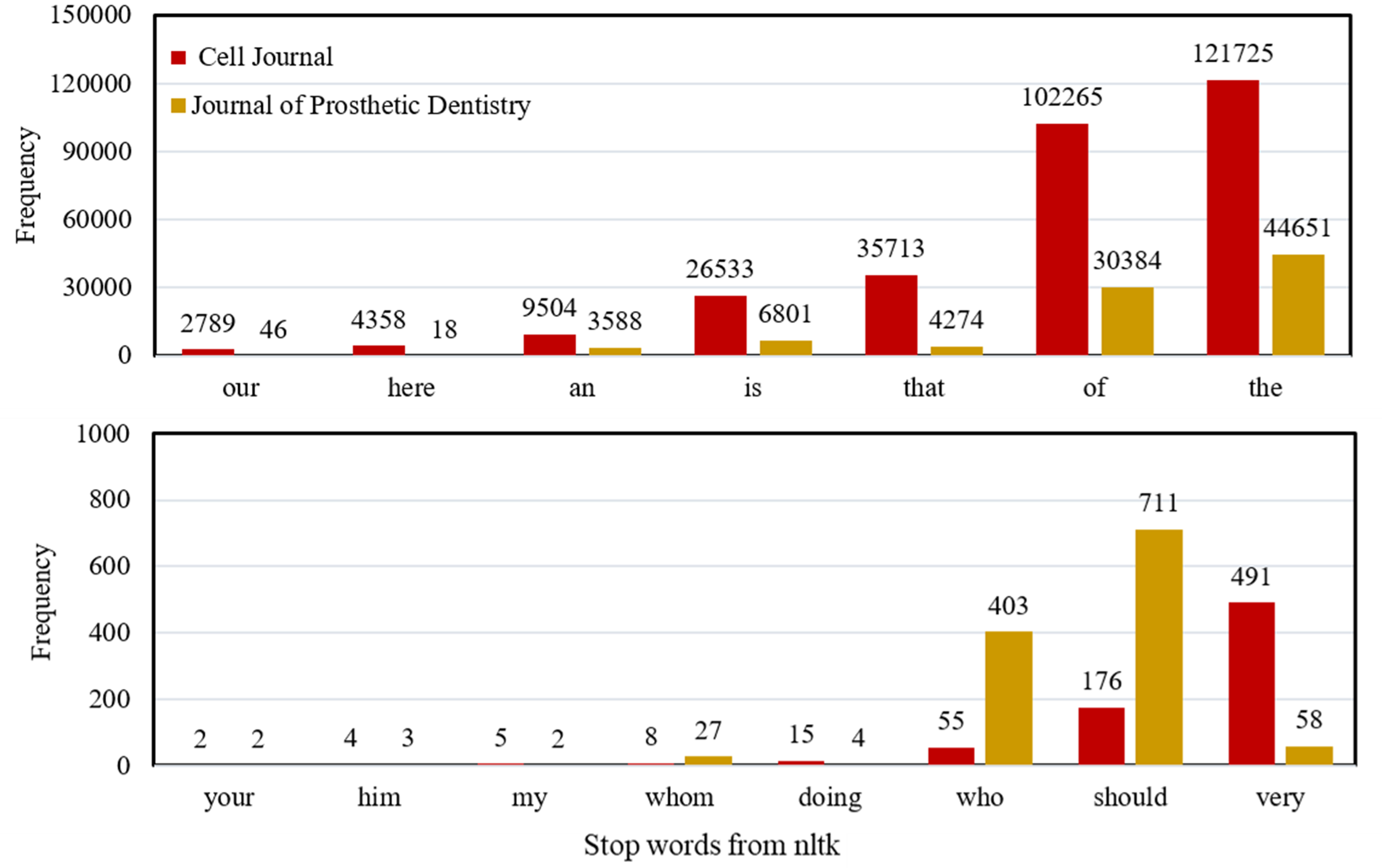}
   \centering
   \caption{nltk stop Words frequency in Cell and Journal of Prosthetic  Dentistry.}
   
   \label{freq}
   \vspace{-.2in}
   \end{figure}

Fig.~\ref{pca} also includes the 127 stop words from the Python nltk  (black stars). It can be noticed that the majority of these words are near the hyperplane, however, few of them defined as longest distance words. To explain such behavior, a sample of nltk stop words which positioned on various distances with respect to the hyperplane is picked as shown in Fig.~\ref{pca}. These words are: ''our'', ``here'', ``an'', ``is'', ``that'', ``of'', ``the'', ``your'', ``him'', ``my'', ``whom'', ``doing'', ``who'', ``should'', and ``very''. Then, the frequency of these words is calculated as shown in Fig.~\ref{freq}. The histogram shows a high discrepancy in the longest distance words’ frequency between the two classes (two journals) such as ``who'', ``should'', and ``whom'', that means these words are most likely to appear in one journal rather than the other.

\medskip
{In this surprising example, some differences are quite significant and indicate that they can be used in determining the classes of papers. Thus, some of them  should not be eliminated by simple traditional stop word removal.}

\subsection{Detect domain-specific words}

Domain-specific words can be defined in this approach as the words with the shortest distance from the hyperplane. Therefore, the final step is to sort the words’ distances in ascending order such that the words that have the shortest distance from the hyperplane can be detected.

\section{Data Sources for Experiments}
We tested our approach on three datasets.  These are chosen to vary in size and complexity for testing the limitations of the approach.

{\it Hansard Speeches 1979--2018:}
This is a public dataset of speeches from the UK House of Commons, extracted from Hansard, the official public records \cite{odell2019}. This data set includes the text for every speech made in the House of Commons between the 1979 general election and the end of 2017, with information for each speaker, including their party affiliation. We reduce the set of speeches to speakers from the two largest parties, Labour and Conservative. The resulting dataset consists of 1,608,012 speeches. We picked the Hansard Speeches because the baseline accuracy of the classifiers applied to these data is much lower than the other datasets, meaning the classification task is more difficult.

\smallskip
{\it IMDB Movie Review Dataset:}
This is a dataset for binary sentiment classification containing 50,000  positive and negative reviews\cite{maas-EtAl:2011:ACL-HLT2011}.  



\smallskip
{\it Pubmed:}
The Pubmed dataset\cite{website2} provides public information by the National Library of  Medicine (NLM). It contains more than 26 million citations for biomedical literature from MEDLINE, life science journals, and online books. Our knowledge base starts as XML files provided by Pubmed, from which we extract each publication abstract, year, and journal. 
In this work we focus on the journal abstracts, treating journal names as the categories into which we categorize the abstracts. To ensure higher quality in the word embeddings, we only consider journals with at least 10,000 abstracts. From these journals, many pairs were generated then tested using the Logistic Regression classifier. The pairs that have an accuracy lower than 98\% were picked to efficiently test the proposed approach. The curated dataset contains 100 pairs of journals and 2,694,790 abstracts.

\section{Validation experiments and results }
We conduct our experiments in two parts: First,  the validation of the proposed approach is studied on the classification accuracy by eliminating three types of words: (1) the words with the shortest distance from the hyperplane (domain-specific words); (2) the words with the longest distance from the hyperplane (distinguishable words); (3) random words. Second, we compare the hyperplane-based approach against two feature selection methods: 
$\chi^2$ and Mutual Information (MI). 

 $\chi^2$ is used as a dimensionality reduction feature selection method which evaluates the independence of the feature and the class. The higher the $\chi^2$ value, the more class information the feature contains. Mutual Information (MI) is another dimensionality reduction feature selection method. MI is used to measure the dependencies between two random variables, which are class and feature. The higher the MI value, the more information content exists between the feature and the class. Accordingly, a low  $\chi^2$ or  MI value indicates that a word is a domain-specific common word. The most distinguishable words are those that have a high $\chi^2$ or MI.  For our proposed hyperplane method, domain-specific words are those that have the shortest distance from the hyperplane.
 
\begin{figure}
\vspace{-.2in}
   \includegraphics[width=0.99\linewidth,height=8cm]{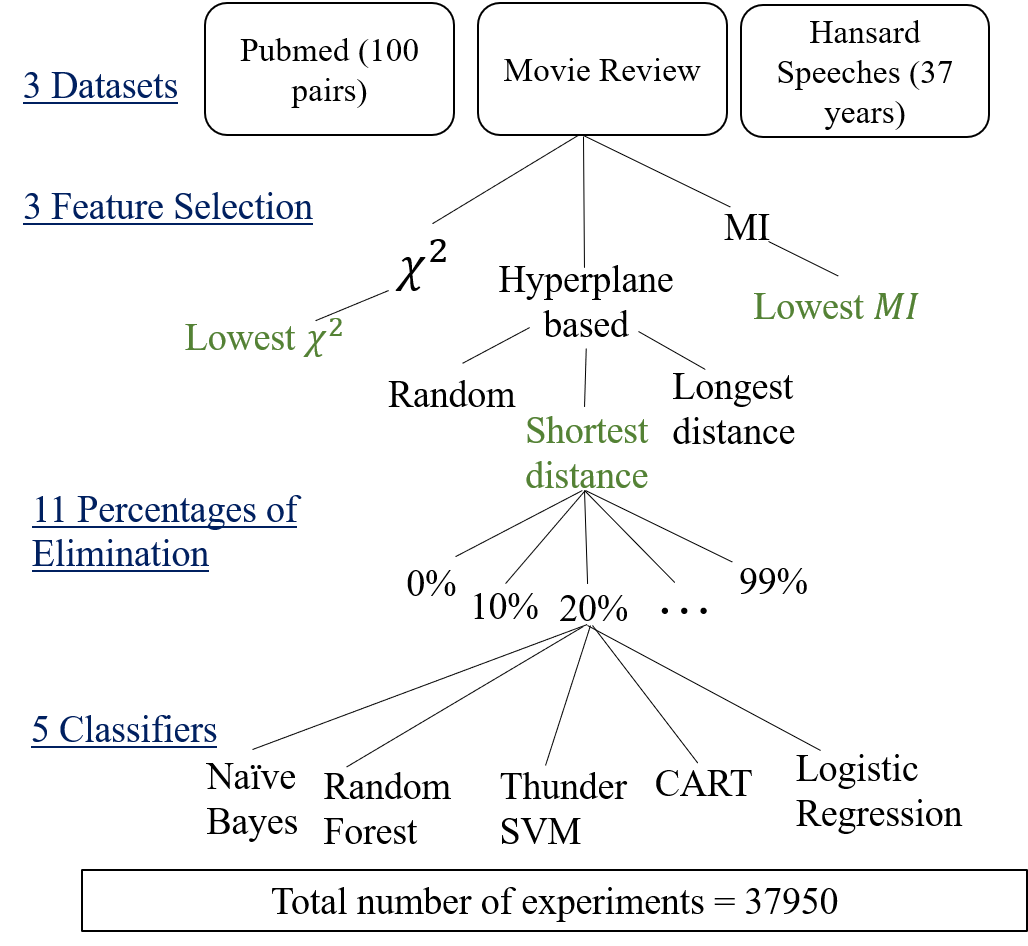}
   \caption{Experimental scheme  }
   \label{exper}
   \vspace{-.2in}
   \end{figure}

Our experimental scheme is illustrated in Fig.~\ref{exper}. In this study, 138 different sub-corpora from the Hansard speeches, IMDB movie review, and  Pubmed datasets are tested per different feature selection methods, percentages of domain-specific words elimination, and classifiers. For each feature selection approach, 11 percentages of domain-specific words are eliminated in  increments of 10 from 0\% to 99\%.  
The 0\% is selected as a reference category because it represents the performance of a classifier before any words are eliminated. 

For each case, five classifiers are applied: Naive Bayes, Thunder SVM, Logistic Regression, Random Forest, and CART. The performance of a classifier is quantified by its prediction accuracy, calculated as the ratio of correctly predicted instances to all instances in the experiment, and estimated using 10-fold cross validation. Overall, we conduct a total of 37,950 experiments to test the hyperplane-based approach. 

\subsection{Hansard Speeches}
\begin{figure}
   \includegraphics[width=0.99\linewidth,height=7cm]{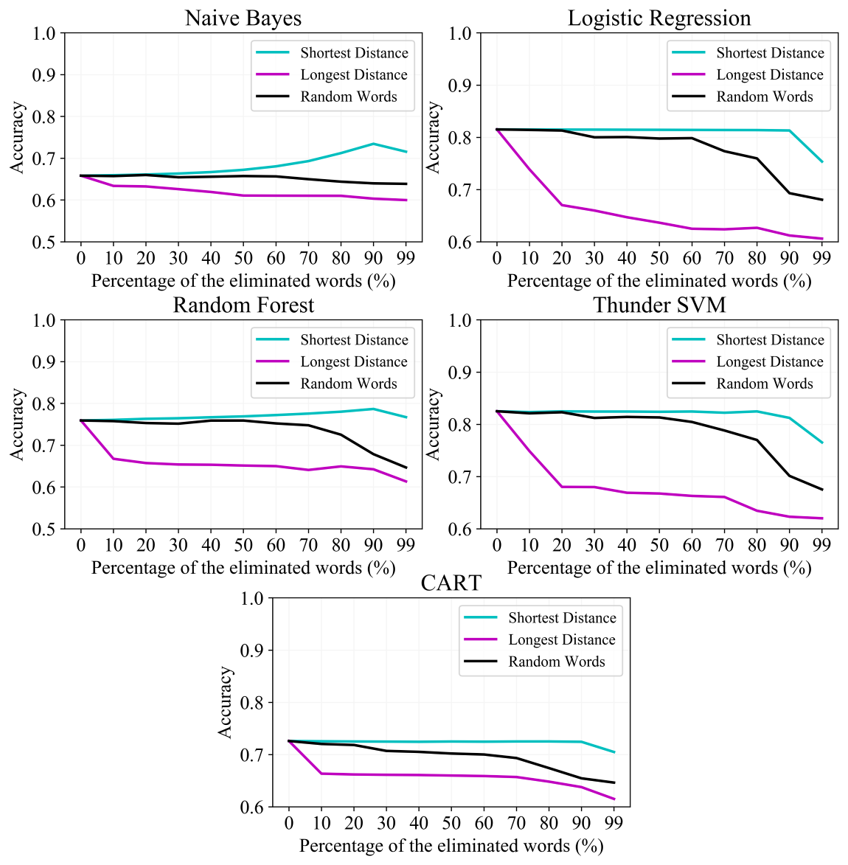}
   \caption{Classifier performance on Hansard Speeches dataset when eliminating different percentages of words based on three different criteria: shortest distance from hyperplane, longest distance from hyperplane, and random word elimination.}
   \label{politics_shr}
   \end{figure}

The Hansard data spans the time period from 1979--2018. We calculate the accuracy for all five classifiers, three feature selection approaches, and 11 elimination percentages for each year. We then calculate average accuracy across all years included in our data. Hyperplane-based approach validation starts with comparing the performance of the classifiers when eliminating three criteria: the shortest distance, longest distance, and random words. Fig.~\ref{politics_shr} presents the average accuracy of classifiers per each selection criterion. As expected the highest accuracy is obtained
when removing the domain-specific words (shortest distance), and the lowest
accuracy when removing the distinguishable words (longest
distance). Also, when the percentage of distinguishable words
elimination increases, accuracy drops significantly. The accuracy of the classifiers when
removing random words is in between the accuracy when removing the shortest or longest distance words. Additionally,
we note that the accuracy of all classifiers when eliminating from 10 to 90\% of words
will maintain about the same level if not increase compared to the reference category (0\% elimination). 

\begin{figure}
\vspace{-.2in}
   \includegraphics[width=0.99\linewidth,height=7cm]{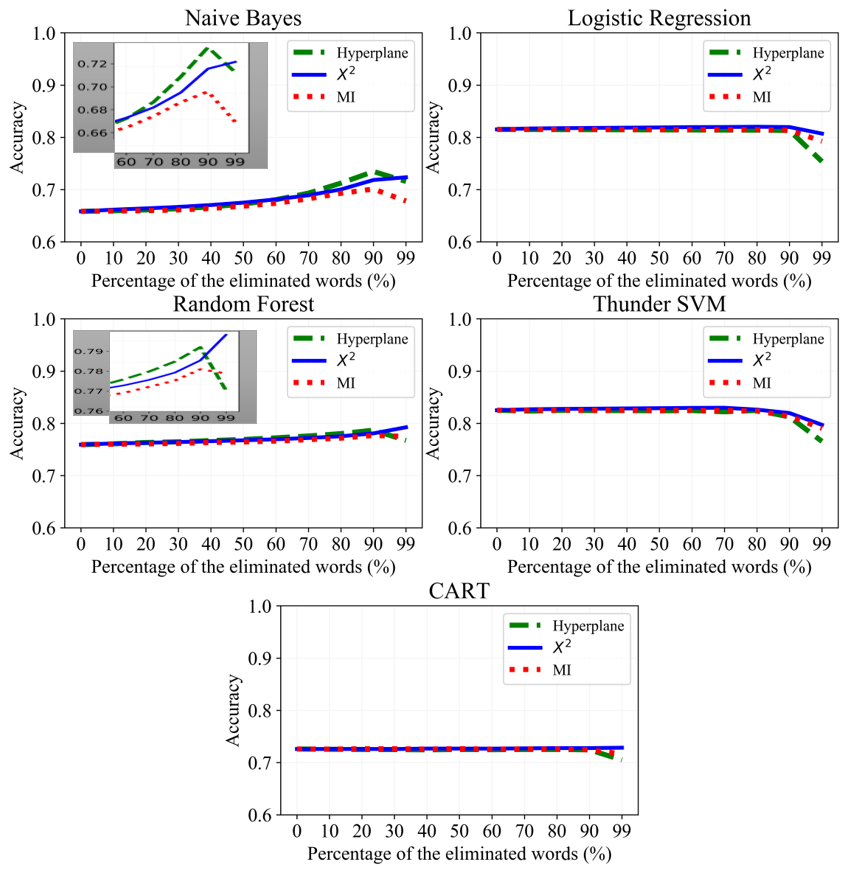}
   \caption{Average classifier accuracy comparison on Hansard Speeches for all the years, using three different approach: hyperplane-based, $\chi^2$ and MI by eliminating different percentages of words starting from shortest to longest distance from hyperplane, lowest to highest $\chi^2$ value and lowest to highest mutual information. }
   \label{speech_comp}
   \end{figure}

Fig.~\ref{speech_comp} presents the average accuracy of the classifiers per hyperplane-based, $\chi^2$ and MI approach on the Hansard Speeches dataset. The hyperplane-based approach improved the average accuracy of the Naive Bayes classifier by 8\% (from 0.66 to 0.74) when eliminating 90\% of domain-specific words with respect to the reference category (0\% elimination). Further, our proposed approach outperforms the $\chi^2$ and MI approaches from 60\% to 90\% elimination. The average accuracy of the Naive Bayes classifier increased at 99\% elimination with respect to the reference experiment in all approaches, where the hyperplane-based approach had the same average accuracy as the  $\chi^2$ and higher than MI approach. The average accuracy of the Random Forest classifier is increased by 2\%, and 3\% when removing 90\% of domain-specific words using our proposed approach,  and $\chi^2$ approach, respectively. MI, in conrast, increased the Random Forest accuracy with less than 1\%. The accuracy of other classifiers (Logistic Regression, Thunder SVM, and CART) stay the same after any elimination percentage. In the case of removing 90\% of domain-specific words, the $\chi^2$ approach and hyperplane-based enhanced the accuracy of both Naive Bayes and Random Forest classifiers compared with the reference experiment.

\begin{figure}
   \includegraphics[width=0.99\linewidth,height=8cm]{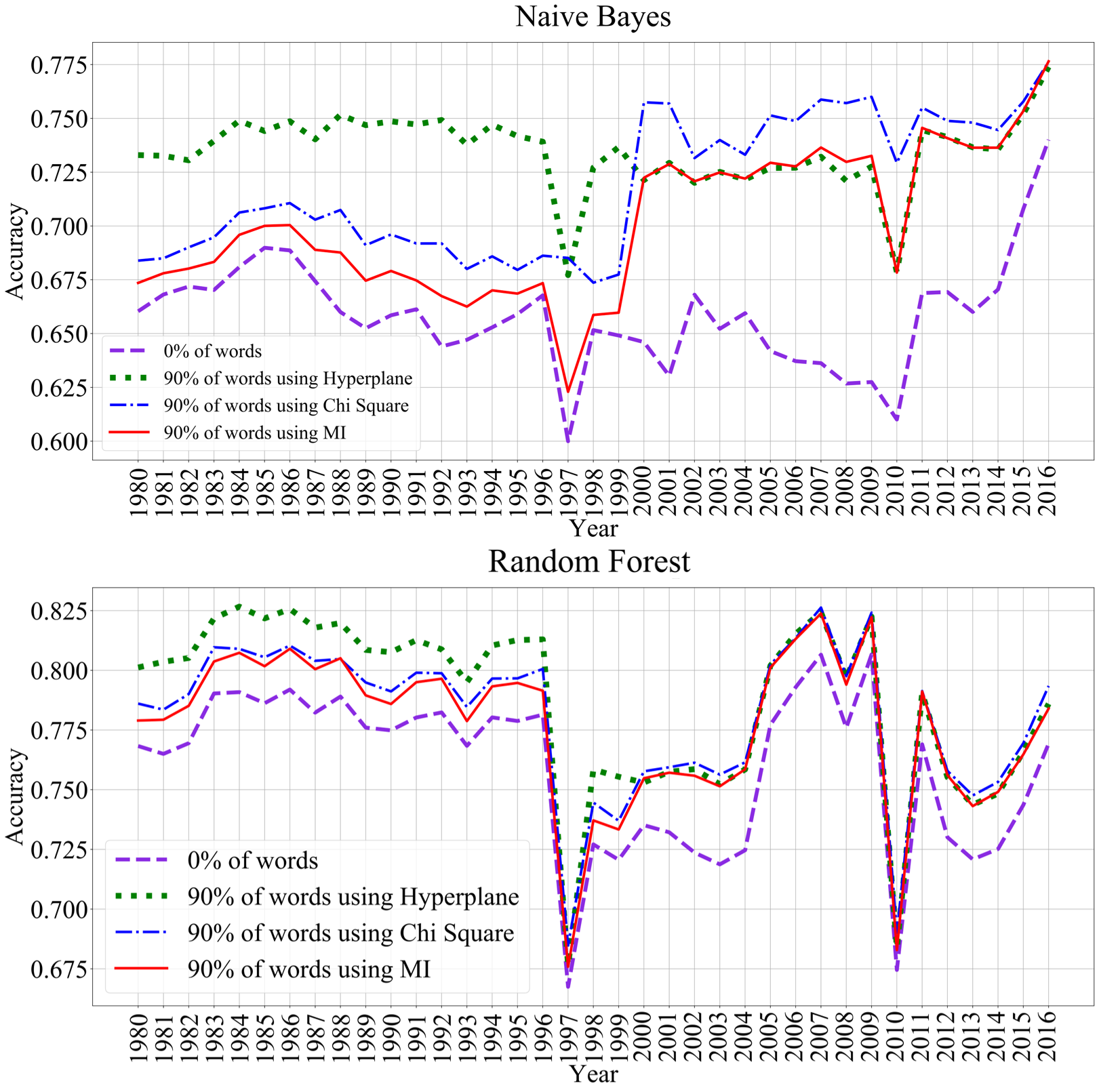}
   \caption{Classifier accuracy on Hansard Speeches dataset per year, using three different approach: hyperplane-based, $\chi^2$, and MI by eliminating 90\%  of words and keeping the words that have longest distance from hyperplane,  highest $\chi^2$ value and highest mutual information, and compare it with classifiers accuracy before any elimination.}
   \label{years}
 \vspace{-.2in} 
   \end{figure}

To further investigate the performance of our proposed hyperplane-based approach, we break the estimation of accuracy down by year instead of calculating the average across years. Fig.~\ref{years} shows trends in accuracy for two classifiers, Naive Bayes and Random Forest, for two elimination categories: 90\% of domain-specific words vs 0\% elimination (our reference category). We notice that the accuracy of both classifiers using our hyperplane-based approach outperform the $\chi^2$ for the years from 1980 to 1999, and outperforms MI for all the years. Our approach further increases the performance of Naive Bayes by 12\% for the year 1992 from 0.64 to 0.76. For Random Forest, our proposed approach increases the accuracy by 4\% in 1981.


Overall, the three approaches play a key role in reducing the dimensionality of the corpus, which means reducing the execution time of the classification problem, and the hyperplane-based approach achieved comparable results with $\chi^2$ and MI. We demonstrate the impact of feature elimination on execution time in Tables~\ref{tab1}. Table~\ref{tab1} presents the average execution time for all years of the five classifiers per elimination percentage.  We can see from this table that when the percentage of elimination increases, the drop in the average execution time increases dramatically. We further compare the execution time for the three elimination strategies hyperplane-based, $\chi^2$, and MI.  
We find that the $\chi^2$ approach has the lowest execution time 8 seconds, while MI has the highest. Our approach outperforms the MI, which only needs 152 seconds to generate the list of words for each year, while MI needs 1743 seconds, more than ten times longer.

\begin{table}[htbp]
\caption{Average Execution Time for each classifier in seconds}\label{tab1}
\begin{center}
\begin{tabular}{|c|c|c|c|c|c|c}
\hline
Percentage of elimination& NB& LR& RF &SVM &CART\\
\hline
0\%&	0.37&	35&	1296&	733&	533\\
\hline
10\%&	0.36&	31&	1260&	686&	525\\
\hline
20\%&	0.35&	28&	1191&	661&	521\\
\hline

30\%&	0.33&	26&	1127&	643&	506\\
\hline
40\%&	0.33&	24&	1099&	626&	502\\
\hline
50\%&	0.31&	22&	1034&	611&	488\\
\hline
60\%&	0.29&	21&	989&	602&	480\\
\hline
70\%&	0.28&	19&	977&	589&	477\\
\hline
80\%&	0.27&	18&	946&	575&	475\\
\hline
90\%&	0.25&	15&	830&	556&	448\\
\hline
99\%&	0.09&	4&	565&	213&	77\\
\hline

\end{tabular}
\end{center}
\end{table}

\begin{figure}
   \includegraphics[width=0.97\linewidth,height=4cm]{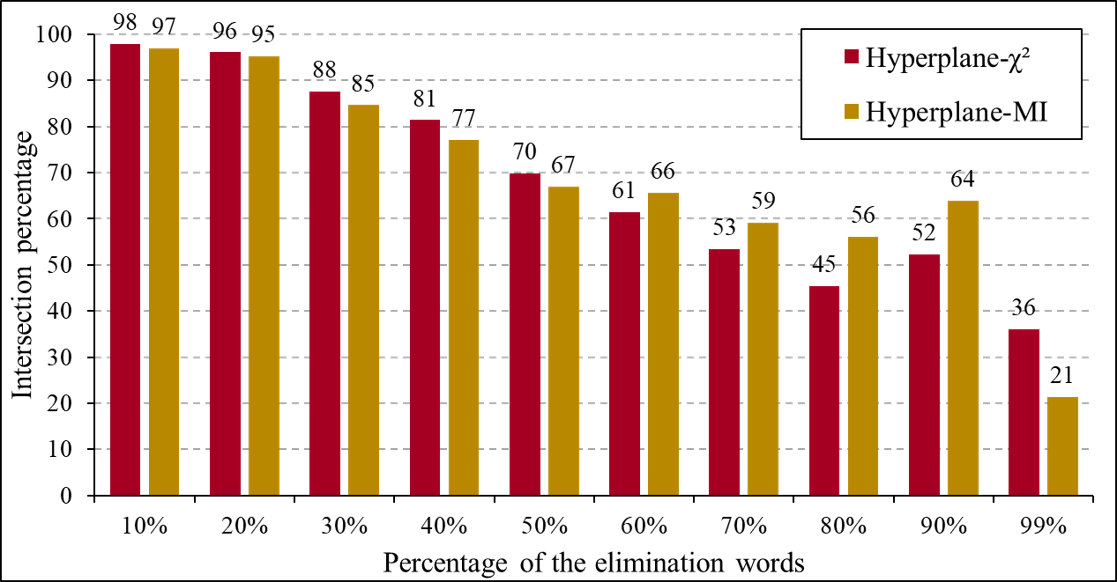}
   \vspace{-.1in}
   \caption{Intersection between Hyperplane-based approach with the $\chi^2$ and MI approach against the range of elimination percentages. 
}
   \label{yearsoverlap}
   \vspace{-.2in}
   \end{figure}

Finally, we investigate the extent to which our approach generates word lists that are different from those generated by the two other approaches. To this end, we look at the intersection of the word sets generated by the three approaches, comparing our method against the other two for different percentage categories of elimination -- see Fig.~\ref{yearsoverlap}. The intersection is defined by the percentage of identical words between two approaches that remain after elimination, divided by the total number of remaining words after elimination.  Fig.~\ref{yearsoverlap} shows that the overlap between the methodologies is not high when eliminate 99\% of words, which means our hyperplane-based approach extracts words that are different from those extracted by the other two approaches, while maintaining the same accuracy. We present sample of the words that used to distinguish between two classes after eliminating 99\% of words in Table~\ref{tab3}. We present examples of the remaining words after eliminating 99\% of words per approach for the year 1980. These are the words with the largest distance from the hyperplane, highest  $\chi^2$, and highest MI.


\begin{table}[htbp]
\vspace{-.02in}
\caption{Presents words from Hansard Speeches dataset for the year 1980 that are deemed most important by the hyperplane, $\chi^2$, and Mutual Information}\label{tab3}
\begin{center}
\begin{tabular}{|c|c|c|}
\hline
Hyperplane words&	$\chi^2$&	MI words\\
\hline
sparkbrook&	minister&	not\\
\hline
disservice&	friend&	for\\
\hline
benchers&	secretary&	hon\\
\hline
dispatch&	state&	and\\
\hline
duchy&	he&	is\\
\hline
sidcup& 	you&	in\\
\hline
engagements&	state&	that\\
\hline
orme&	conservative&	to\\
\hline
mislead&	she&	of\\
\hline
\end{tabular}
\end{center}
\vspace{-.1in}
\end{table}

 

\subsection{IMDB Movie Review Dataset}
\begin{figure}
   \includegraphics[width=0.99\linewidth,height=8cm]{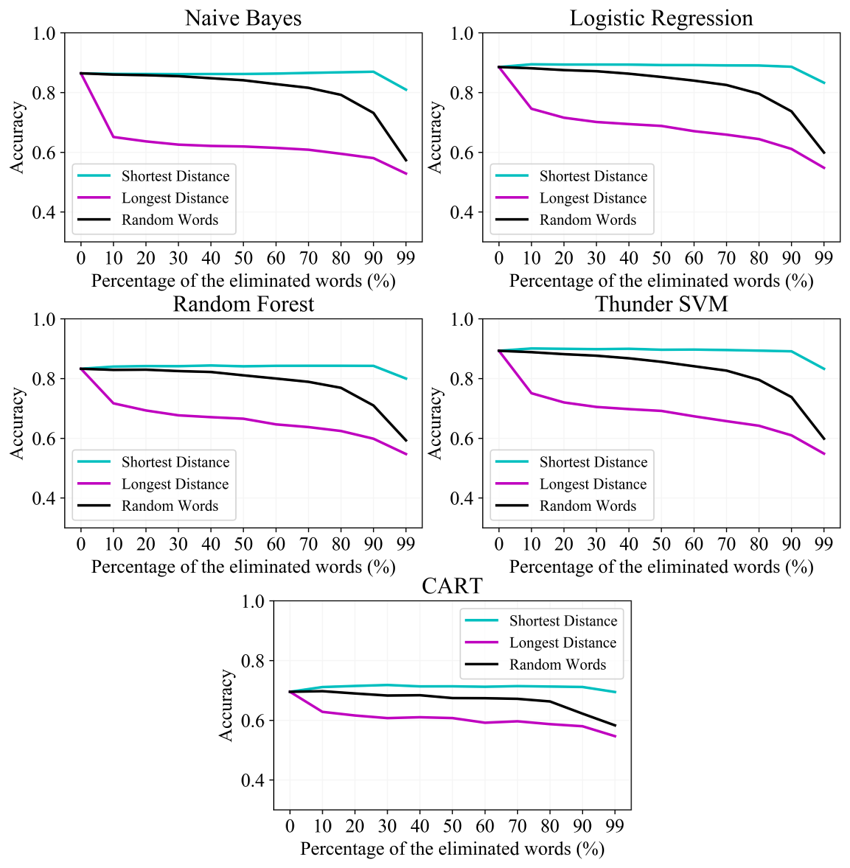}
   \vspace{-.1in}
   \caption{Classifiers performance on IMDB movie review dataset by eliminating different percentage of words that have shortest, longest distance from hyperplane and by eliminating random words. }
   \vspace{-.1in}
   \label{movie_SHR}
   \end{figure}
\begin{figure}
\centering
   \includegraphics[width=0.99\linewidth,height=8cm]{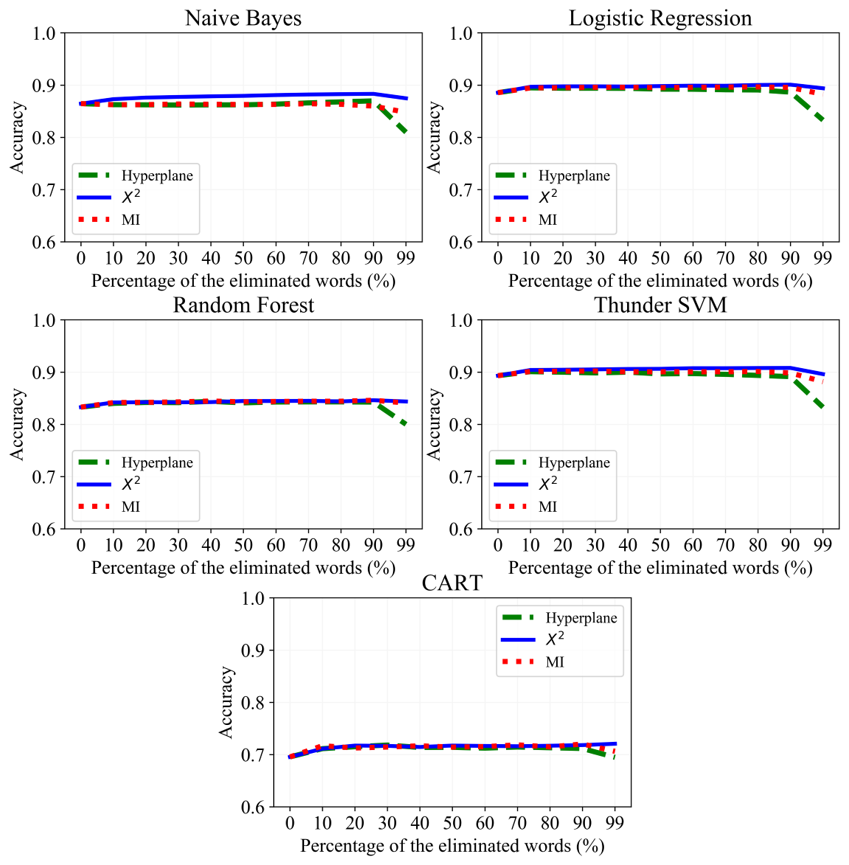}
  %
   \caption{Classifier performance comparison on IMDB movie review dataset using three different approach: hyperplane-based, $\chi^2$ and MI by eliminating different percentages of words starting from shortest to longest distance from hyperplane, lowest to highest $\chi^2$ value and lowest to highest mutual information.  }
   \label{movie_comp}
   \end{figure}

The accuracy
of the five classifiers with respect to the range elimination
percentages for the shortest distance, longest distance, and random words estimated from this dataset show the same trend as the Hansard Speeches dataset, as is illustrated in Fig.~\ref{movie_SHR}. 
Fig.~\ref{movie_comp} shows the classifier performance of testing the IMDB movie review dataset for the hyperplane-based approach and the two feature selection methods using five classifiers. It can be noticed that the  accuracy for the Naive Bayes and Thunder SVM classifiers are improved by using the three approaches when eliminating 90\% of words. The hyperplane-based approach performs better than the MI approach when eliminating 90\% of the domain-specific words under Naive Bayes. The $\chi^2$ is slightly higher than the hyperplane-based approach and the MI at all elimination percentages under Naive Bayes. The accuracy of the other classifiers (Logistic Regression, Random Forest, and CART) is stable before and after eliminating 90\% of words using all approaches. 

When removing 99\% of domain-specific words, only $\chi^2$ maintains the same accuracy, while the performance of the other two approaches drops. We note that the drop in performance cannot be explained by the small number of data points at this elimination level alone. After eliminating 99\% of the words, there are 2,221 words remaining, with an average of 45-80 words per review, depending on which approach is being used, which should be enough distinguishable words to classify the data. As such, the drop in performance is a shortcoming of the two methods, though only at this high level of word elimination. 

\begin{figure}
   \includegraphics[width=0.97\linewidth,height=4cm]{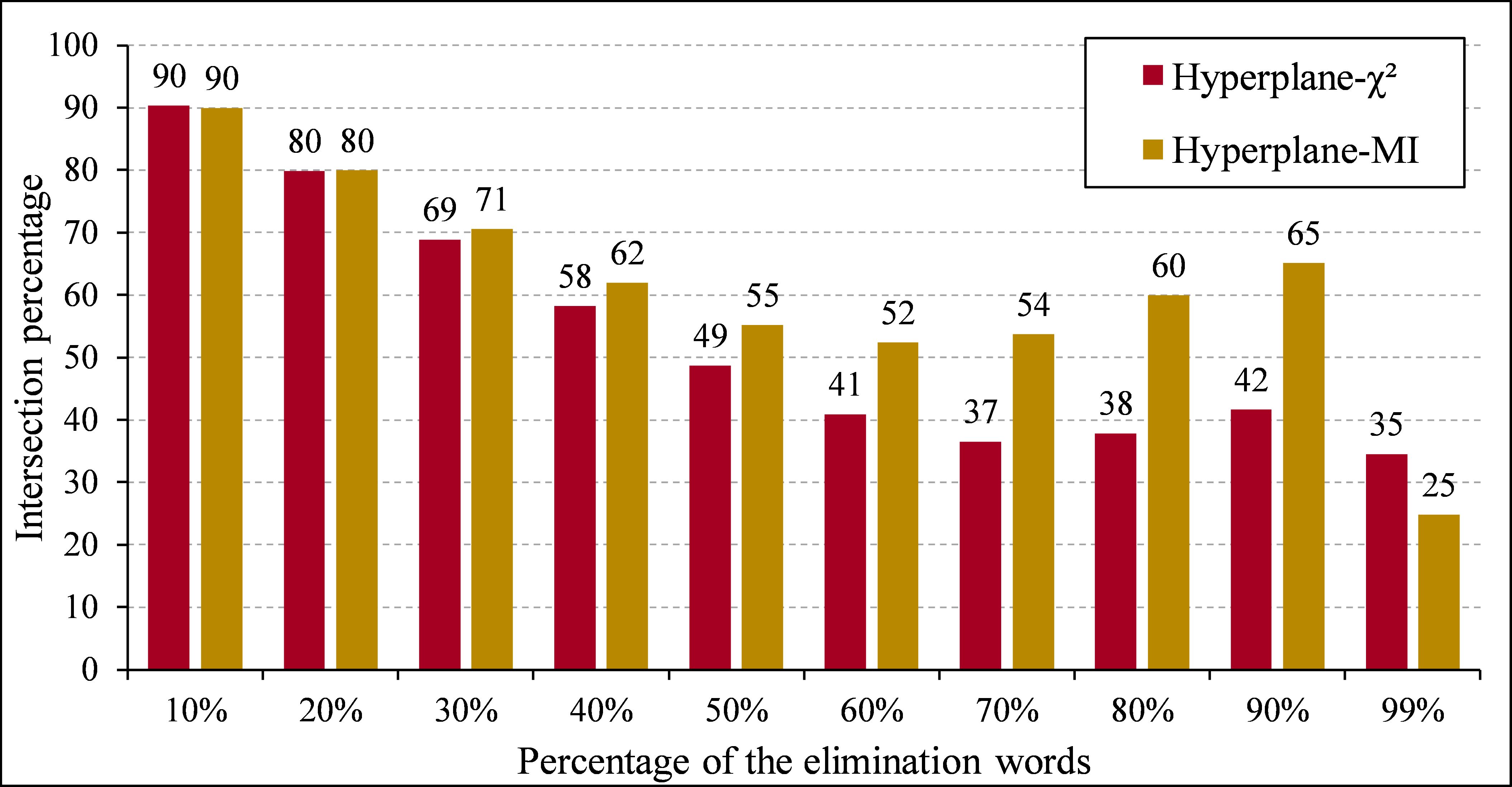}
   \caption{Intersection between Hyperplane-based approach with the $\chi^2$ and MI approach against the range of elimination percentages. 
}
   \label{movie_overlap}
   \vspace{-.2in}
   \end{figure}

The intersection of the remaining words between the proposed approach with the $\chi^2$ and MI approach against the range of elimination show the same trend as we observed in the Hansard Speeches dataset, which is illustrated in Fig.~\ref{movie_overlap}. We also again present examples of the remaining words after eliminating 99\% in Table~\ref{tab4}. 



\begin{table}[htbp]
\caption{Presents words from IMDB Movie Review Dataset that are deemed most important by the hyperplane, $\chi^2$, and Mutual Information}\label{tab4}
\begin{center}
\begin{tabular}{|c|c|c|}
\hline
Hyperplane words&	$\chi^2$ words&	MI words\\
\hline
waste&	bad&	the \\
\hline
renting&	worst&	and \\
\hline
crap&	waste&	of\\
\hline
stupid&	awful&	to\\
\hline
suck&	great&	this\\
\hline
bother&	terrible&	is\\
\hline
unwatchable&	horrible&	in\\
\hline
pile&	excellent&	It\\
\hline
\end{tabular}
\end{center}
\vspace{-.1in}
\end{table}

\subsection{Pubmed}
\begin{figure}
   \includegraphics[width=0.99\linewidth,height=8cm]{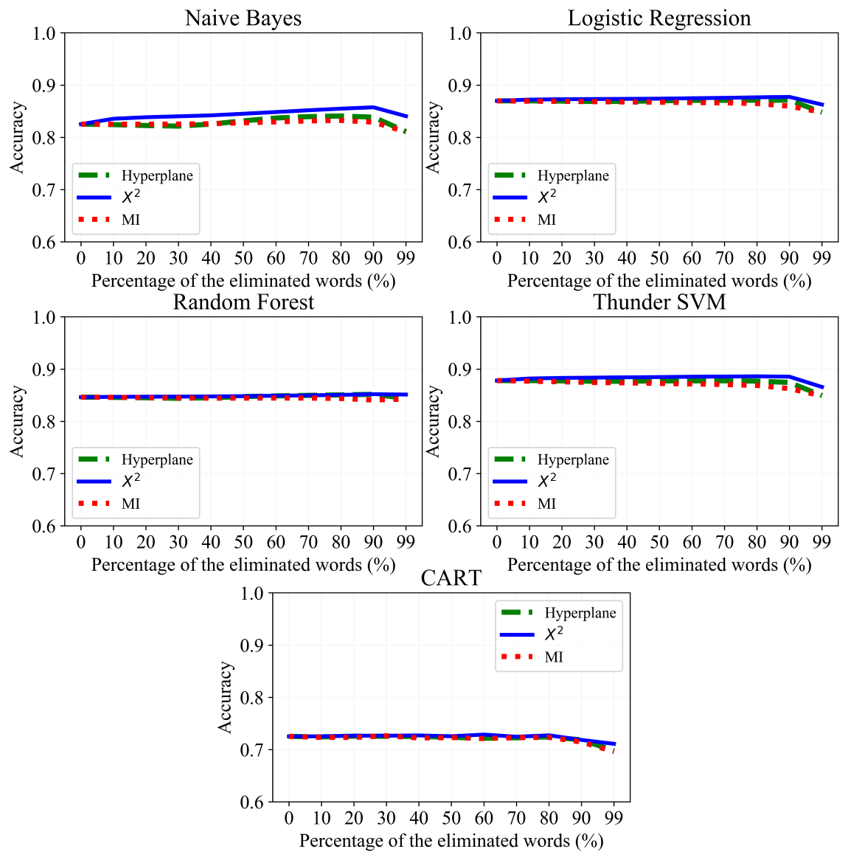}
   \caption{Classifier performance comparison on 100 pairs of Pubmed datasets  using three different approach: hyperplane-based, $\chi^2$ and MI by eliminating different percentages of words starting from shortest to longest distance from hyperplane, lowest to highest $\chi^2$ value and lowest to highest mutual information  }
   \label{100pair}
   \end{figure}

The curated Pubmed  dataset includes abstracts from 100 pairs of different journals. For each pair, the accuracy of all classifiers is calculated per all elimination percentages. Then, the average accuracy for all pairs is calculated and presented. Fig.~\ref{100pair} presents the average accuracy
of the five classifiers with respect to the range of elimination
percentages for the shortest distance, longest distance, and random words to the Pubmed dataset  
have the same trend to  the Hansard  Speeches dataset and the IMDB movie review  dataset.

\begin{figure}
 \includegraphics[width=0.99\linewidth,height=7cm]{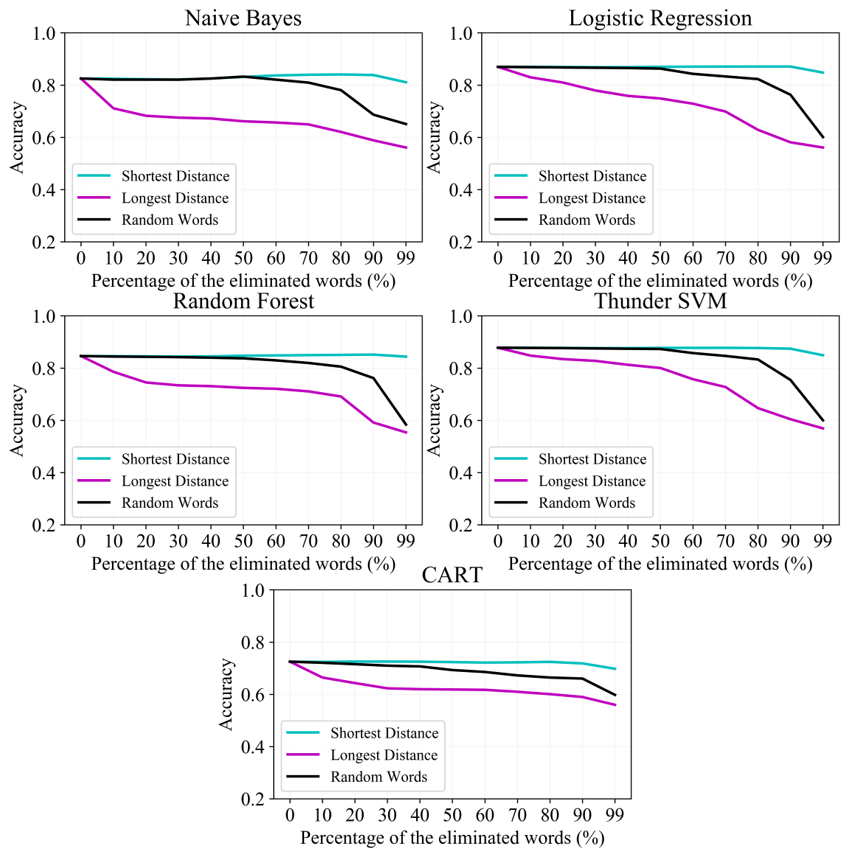}
 \caption{Classifiers performance on 100 pairs of Pubmed dataset by eliminating different percentage of words that have shortest, longest distance from hyperplane and by eliminating random words }
   \label{100pair_SRH}
   \end{figure}

Fig.~\ref{100pair}  shows the average classifier accuracy of testing the 100 pairs of Pubmed dataset for the hyperplane-based and two feature selection methods and five classifiers. The classifying accuracy for the Naive Bayes is improved by using the hyperplane-based and $\chi^2$. The hyperplane-based approach outperforms the MI approach when eliminating 90\% of the domain-specific words for Naive Bayes, Logistic Regression and Thunder SVM. The accuracy of the other classifiers (Random Forest, and CART)  is stable before and after eliminating 90\% of words using all approaches. 
 
\begin{figure}
   \includegraphics[width=0.97\linewidth,height=4cm]{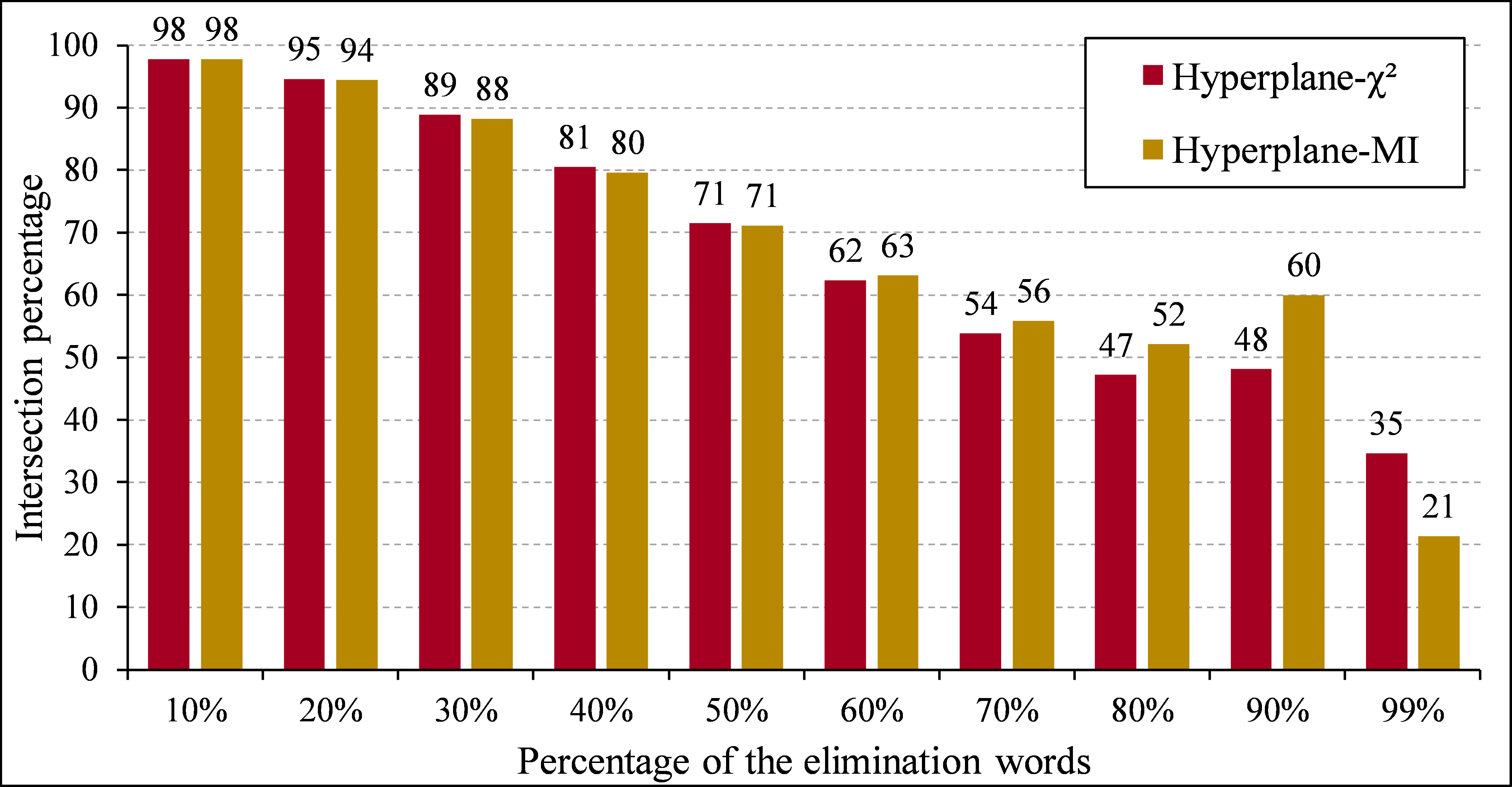}
   \caption{Intersection between Hyperplane-based aproach with the $\chi^2$ and MI approach against the range of elimination percentages.
}
   \label{pubmedoverlap}
   \vspace{-.2in}
   \end{figure}

 We present the intersection of the remaining words between the proposed approach with the $\chi^2$ and MI approach against the range of elimination on one pair, Cell and Journal of Prosthetic Dentistry, in Fig.~\ref{pubmedoverlap}. We find that the intersections have the same trend as those observed on the two other datasets. We present a sample of the remaining words after eliminating 99\% of words in Table~\ref{tab5}. 



\begin{table}[htbp]
 \vspace{-.2in}
\caption{Presents words of the Cell and Journal of Prosthetic Dentistry from  Pubmed  dataset that are deemed most important by the hyperplane, $\chi^2$, and Mutual Information }\label{tab5}
\begin{center}
\begin{tabular}{|c|c|c|}
\hline
Hyper plane words&	$\chi^2$ words&	MI words\\

\hline
 computeraided& 	denture&the	\\
\hline
 restorations& 	cell&of	\\ 
\hline
provisional & 	we&and	\\
\hline
castings & 	dental&in	\\
\hline
impressions & 	implant&to	\\
\hline
crowns& 	resin&that	\\
\hline
fabrication & 	teeth&is	\\
\hline
\end{tabular}
\end{center}
\vspace{-.2in}
\end{table}

\section{Conclusion}
This study proposed a novel mathematical approach for detecting domain-specific words called the Hyperplane-based approach. This new approach depends on the notion of low dimensional representation of the word in vector space and its distance from the hyperplane, where the domain-specific words are defined as the words with the shortest distance from the separating hyperplane. The performance of the proposed approach is quantified by the accuracy and the execution time of five classifiers: (1) Naive Bayes, (2) Random Forest, (3) Logistic Regression, (4) Thunder SVM, and (5) CART. This approach is validated using 138 sub-corpora from three datasets (Hansard Speeches, IMDB Movie Review, and Pubmed). Also, it is compared with two feature selection approaches, namely $\chi^2$ and MI. For each feature selection method including the hyperplane-based approach and each corpus, various word elimination percentages are considered to find the optimal elimination percentage for each classifier and approach. 
The hyperplane-based approach generally improves the performance of the classifier and it achieved comparable performance with the $\chi^2$ and MI. \emph{However, our experiments indicate that qualitatively the eliminated words significantly differ from other approaches. In addition, the method is more robust to the erroneous elimination of important words.} The performance of our approach varies with respect to the classifier and the elimination percentages. For example, the Naive Bayes classifier presented the best improvement of accuracy before and after the elimination using our approach, and the optimal elimination percentage per our approach is 90\% for all datasets.
Finally, the proposed approach plays a key role in reducing the dimensionality of the corpus, which means reducing the  classification execution time. The implementation of the hyperplane-based approach in different datasets is straight-forward and merging the hyperplane based-domain words with other feature selection domain words is a future research recommendation. Replicability: Our code and results will be made available online when the paper is published. 

\medskip
\noindent
\textsc{Acknowledgements:}
This research has been supported by NSF grants \#1405767, \#2027864 and \#1725573.

\bibliographystyle{IEEEtran}
\bibliography{main}

\begin{thebibliography}{10}
\providecommand{\url}[1]{#1}
\csname url@samestyle\endcsname
\providecommand{\newblock}{\relax}
\providecommand{\bibinfo}[2]{#2}
\providecommand{\BIBentrySTDinterwordspacing}{\spaceskip=0pt\relax}
\providecommand{\BIBentryALTinterwordstretchfactor}{4}
\providecommand{\BIBentryALTinterwordspacing}{\spaceskip=\fontdimen2\font plus
\BIBentryALTinterwordstretchfactor\fontdimen3\font minus
  \fontdimen4\font\relax}
\providecommand{\BIBforeignlanguage}[2]{{%
\expandafter\ifx\csname l@#1\endcsname\relax
\typeout{** WARNING: IEEEtran.bst: No hyphenation pattern has been}%
\typeout{** loaded for the language `#1'. Using the pattern for}%
\typeout{** the default language instead.}%
\else
\language=\csname l@#1\endcsname
\fi
#2}}
\providecommand{\BIBdecl}{\relax}
\BIBdecl

\bibitem{zou2006automatic}
F.~Zou, F.~L. Wang, X.~Deng, S.~Han, and L.~S. Wang, ``Automatic construction
  of chinese stop word list,'' in \emph{Proceedings of the 5th WSEAS
  international conference on Applied computer science}, 2006, pp. 1010--1015.

\bibitem{chong2012empirical}
T.~Y. Chong, R.~E. Banchs, and E.~S. Chng, ``An empirical evaluation of stop
  word removal in statistical machine translation,'' in \emph{Proceedings of
  the Joint Workshop on Exploiting Synergies between Information Retrieval and
  Machine Translation (ESIRMT) and Hybrid Approaches to Machine Translation
  (HyTra)}, 2012, pp. 30--37.

\bibitem{jha2016hsra}
V.~Jha, N.~Manjunath, P.~D. Shenoy, and K.~Venugopal, ``Hsra: Hindi stopword
  removal algorithm,'' in \emph{2016 international conference on
  microelectronics, computing and communications (MicroCom)}.\hskip 1em plus
  0.5em minus 0.4em\relax IEEE, 2016, pp. 1--5.

\bibitem{bell1990text}
T.~C. Bell, J.~G. Cleary, and I.~H. Witten, \emph{Text compression}.\hskip 1em
  plus 0.5em minus 0.4em\relax Prentice-Hall, Inc., 1990.

\bibitem{avudaiappan2017detecting}
N.~Avudaiappan, A.~Herzog, S.~Kadam, Y.~Du, J.~Thatche, and I.~Safro,
  ``Detecting and summarizing emergent events in microblogs and social media
  streams by dynamic centralities,'' in \emph{2017 IEEE International
  Conference on Big Data (Big Data)}.\hskip 1em plus 0.5em minus 0.4em\relax
  IEEE, 2017, pp. 1627--1634.

\bibitem{9005964}
C.~{Gropp}, A.~{Herzog}, I.~{Safro}, P.~W. {Wilson}, and A.~W. {Apon},
  ``Clustered latent {D}irichlet allocation for scientific discovery,'' in
  \emph{2019 IEEE International Conference on Big Data (Big Data)}, 2019, pp.
  4503--4511.

\bibitem{website2}
2016. PubMed. (2016). \url{https://www.ncbi.nlm.nih.gov/pubmed/}.

\bibitem{maas-EtAl:2011:ACL-HLT2011}
\BIBentryALTinterwordspacing
A.~L. Maas, R.~E. Daly, P.~T. Pham, D.~Huang, A.~Y. Ng, and C.~Potts,
  ``Learning word vectors for sentiment analysis,'' in \emph{Proceedings of the
  49th Annual Meeting of the Association for Computational Linguistics: Human
  Language Technologies}.\hskip 1em plus 0.5em minus 0.4em\relax Portland,
  Oregon, USA: Association for Computational Linguistics, June 2011, pp.
  142--150. [Online]. Available: \url{http://www.aclweb.org/anthology/P11-1015}
\BIBentrySTDinterwordspacing

\bibitem{odell2019}
\BIBentryALTinterwordspacing
E.~Odell, ``Hansard {{Speeches V2}}.6.0 [dataset].'' [Online]. Available:
  \url{https://evanodell.com/projects/datasets/hansard-data/}
\BIBentrySTDinterwordspacing

\bibitem{kaur2018systematic}
J.~Kaur and P.~K. Buttar, ``A systematic review on stopword removal
  algorithms,'' \emph{Int. J. Futur. Revolut. Comput. Sci. Commun. Eng},
  vol.~4, no.~4, 2018.

\bibitem{1241221}
M.~P. {Sinka} and D.~W. {Corne}, ``Towards modernised and web-specific
  stoplists for web document analysis,'' in \emph{Proceedings IEEE/WIC
  International Conference on Web Intelligence (WI 2003)}, 2003, pp. 396--402.

\bibitem{garg2014effect}
U.~Garg, V.~Goyal, U.~Garg, and V.~Goyal, ``Effect of stop word removal on
  document similarity for hindi text,'' \emph{An Int. Jounal Eng. Sci.},
  vol.~2, no. December, 2014.

\bibitem{gupta2011preprocessing}
V.~Gupta and G.~S. Lehal, ``Preprocessing phase of punjabi language text
  summarization,'' in \emph{International Conference on Information Systems for
  Indian Languages}.\hskip 1em plus 0.5em minus 0.4em\relax Springer, 2011, pp.
  250--253.

\bibitem{puri2013automated}
R.~Puri, R.~Bedi, and V.~Goyal, ``Automated stopwords identification in punjabi
  documents,'' \emph{An Int. J. Eng. Sci.}, vol.~8, no. June, pp. 119--125,
  2013.

\bibitem{zheng2011comparative}
G.~Zheng and G.~Gaowa, ``A comparative study on between mongolian stop words
  and english stop words,'' \emph{Journal of chinese information processing},
  vol.~4, pp. 35--38, 2011.

\bibitem{rani2018automatic}
R.~Rani and D.~Lobiyal, ``Automatic construction of generic stop words list for
  hindi text,'' \emph{Procedia computer science}, vol. 132, pp. 362--370, 2018.

\bibitem{miretie2018automatic}
S.~G. Miretie and V.~Khedkar, ``Automatic generation of stopwords in the
  amharic text,'' \emph{International Journal of Computer Applications}, vol.
  975, p. 8887, 2018.

\bibitem{zou2006stop}
F.~Zou, F.~L. Wang, X.~Deng, S.~Han, and L.~S. Wang, ``Stop word list
  construction and application in chinese language processing,'' \emph{WSEAS
  Transactions on Information Science and Applications}, vol.~3, no.~6, pp.
  1036--1044, 2006.

\bibitem{alajmi2012toward}
A.~Alajmi, E.~Saad, and R.~Darwish, ``Toward an arabic stop-words list
  generation,'' \emph{International Journal of Computer Applications}, vol.~46,
  no.~8, pp. 8--13, 2012.

\bibitem{lo2005automatically}
R.~T.-W. Lo, B.~He, and I.~Ounis, ``Automatically building a stopword list for
  an information retrieval system,'' in \emph{Journal on Digital Information
  Management: Special Issue on the 5th Dutch-Belgian Information Retrieval
  Workshop (DIR)}, vol.~5, 2005, pp. 17--24.

\bibitem{raulji2016stop}
J.~K. Raulji and J.~R. Saini, ``Stop-word removal algorithm and its
  implementation for sanskrit language,'' \emph{International Journal of
  Computer Applications}, vol. 150, no.~2, pp. 15--17, 2016.

\bibitem{rakholia2017rule}
R.~M. Rakholia and J.~R. Saini, ``A rule-based approach to identify stop words
  for gujarati language,'' in \emph{Proceedings of the 5th International
  Conference on Frontiers in Intelligent Computing: Theory and
  Applications}.\hskip 1em plus 0.5em minus 0.4em\relax Springer, 2017, pp.
  797--806.

\bibitem{4721850}
L.~{Hao} and L.~{Hao}, ``Automatic identification of stop words in chinese text
  classification,'' in \emph{2008 International Conference on Computer Science
  and Software Engineering}, vol.~1, 2008, pp. 718--722.

\bibitem{mikolov2013linguistic}
T.~Mikolov, W.-t. Yih, and G.~Zweig, ``Linguistic regularities in continuous
  space word representations,'' in \emph{Proceedings of the 2013 conference of
  the north american chapter of the association for computational linguistics:
  Human language technologies}, 2013, pp. 746--751.

\bibitem{lewis1998naive}
D.~D. Lewis, ``Naive (bayes) at forty: The independence assumption in
  information retrieval,'' in \emph{European conference on machine
  learning}.\hskip 1em plus 0.5em minus 0.4em\relax Springer, 1998, pp. 4--15.

\bibitem{sadrfaridpour2019engineering}
E.~Sadrfaridpour, T.~Razzaghi, and I.~Safro, ``Engineering fast multilevel
  support vector machines,'' \emph{Machine Learning}, vol. 108, no.~11, pp.
  1879--1917, 2019.

\bibitem{wenthundersvm18}
Z.~Wen, J.~Shi, Q.~Li, B.~He, and J.~Chen, ``{ThunderSVM}: A fast {SVM} library
  on {GPUs} and {CPUs},'' \emph{Journal of Machine Learning Research}, vol.~19,
  pp. 797--801, 2018.

\bibitem{lever2016logistic}
J.~Lever, M.~Krzywinski, and N.~Altman, ``Logistic regression,'' 2016.

\bibitem{ho1995random}
T.~K. Ho, ``Random decision forests,'' in \emph{Proceedings of 3rd
  international conference on document analysis and recognition}, vol.~1.\hskip
  1em plus 0.5em minus 0.4em\relax IEEE, 1995, pp. 278--282.

\bibitem{breiman1984classification}
L.~Breiman, J.~Friedman, C.~J. Stone, and R.~A. Olshen, \emph{Classification
  and regression trees}.\hskip 1em plus 0.5em minus 0.4em\relax CRC press,
  1984.

\bibitem{scikit-learn}
F.~Pedregosa, G.~Varoquaux, A.~Gramfort, V.~Michel, B.~Thirion, O.~Grisel,
  M.~Blondel, P.~Prettenhofer, R.~Weiss, V.~Dubourg, J.~Vanderplas, A.~Passos,
  D.~Cournapeau, M.~Brucher, M.~Perrot, and E.~Duchesnay, ``Scikit-learn:
  Machine learning in {P}ython,'' \emph{Journal of Machine Learning Research},
  vol.~12, pp. 2825--2830, 2011.

\bibitem{mikolov2013efficient}
T.~Mikolov, K.~Chen, G.~Corrado, and J.~Dean, ``Efficient estimation of word
  representations in vector space,'' \emph{arXiv preprint arXiv:1301.3781},
  2013.

\end{thebibliography}
\vspace{12pt}
\color{red}




\end{document}